\documentclass[11pt,draftcls,a4paper,onecolumn]{IEEEtran}%\usepackage{endfloat}

\usepackage{amsmath,epsfig,amssymb,verbatim,amsopn,citesort,subfigure,color,multirow}

\renewcommand{\baselinestretch}{1.9}

\newcommand{\ve}[1]{\boldsymbol{#1}}

\newcommand{\vA}{\ve{A}} \newcommand{\va}{\ve{a}}
\newcommand{\vB}{\ve{B}} \newcommand{\vb}{\ve{b}}

\newcommand{\vG}{\ve{G}} \newcommand{\vg}{\ve{g}}
\newcommand{\vH}{\ve{H}} \newcommand{\vh}{\ve{h}}
\newcommand{\vI}{\ve{I}}

\newcommand{\vN}{\ve{N}} \newcommand{\vn}{\ve{n}}
 
\newcommand{\vP}{\ve{P}} \newcommand{\vp}{\ve{p}}
\newcommand{\vQ}{\ve{Q}} 
\newcommand{\vR}{\ve{R}}

\newcommand{\vU}{\ve{U}}

\newcommand{\vX}{\ve{X}} \newcommand{\vx}{\ve{x}}
\newcommand{\vY}{\ve{Y}} \newcommand{\vy}{\ve{y}}
\newcommand{\vZ}{\ve{Z}}

\newcommand{\vlam}{\ve{\lambda}}
\newcommand{\vSig}{\ve{\Sigma}}

\newcommand{\con}[1]{{#1}^{\ast}}
\newcommand{\mct}[1]{{#1}^{\dagger}}
\newcommand{\mt}[1]{{#1}^{T}}
\newcommand{\RxA}{N_r}
\newcommand{\TxA}{N_t}

\newcommand{\Ld}{L_d}
\newcommand{\Lp}{L_p}
\newcommand{\LpOpt}{L^*_{p}}
\newcommand{\Ts}{T_s}

\newcommand{\Pm}{\mathcal{P}}
\newcommand{\Pd}{\mathcal{P}_d}
\newcommand{\Pda}{\mathcal{P}_{d,1}}
\newcommand{\Pdb}{\mathcal{P}_{d,2}}
\newcommand{\Pp}{\mathcal{P}_p}

\newcommand{\SNReff}{\rho_{\text{eff}}}
\newcommand{\SNReffi}{\rho_{\text{eff},i}}

\newcommand{\Rh}{\vR_{\vH}}
\newcommand{\Rhhat}{\vR_{\hhat}}
\newcommand{\Rhtilde}{\vR_{\htilde}}

\newcommand{\htilde}{\tilde{\vH}}
\newcommand{\hhat}{\hat{\vH}}

\newcommand{\sn}{\sigma^2_{\vn}}

\newcommand{\shhat}{\sigma^2_{\hhat}}
\newcommand{\shtilde}{\sigma^2_{\htilde}}

\newcommand{\Clb}{\overline{C}_{\rm{LB}}}
\newcommand{\Clbl}{C_{{\rm{LB}}}}
\newcommand{\Cubl}{C_{{\rm{UB}}}}
\newcommand{\Cgap}{C_{\rm{gap}}}

\def\onedot{.,\,}
\def\eg{\emph{e.g}\onedot} 

\def\ie{\emph{i.e}\onedot}

\def\wrt{w.r.t\onedot}

\newtheorem{Theorem}{Theorem}

\newtheorem{Definition}{Definition}

\newtheorem{Corollary}{Corollary}

\newcommand{\AuthorOne}{Xiangyun~Zhou}
\newcommand{\AuthorTwo}{Parastoo~Sadeghi}
\newcommand{\AuthorThree}{Tharaka~A.~Lamahewa}
\newcommand{\AuthorFour}{Salman~Durrani}

\newcommand{\ThankOne}{This paper was presented in part at the 2008 IEEE International Conference
on Communication Systems, Guangzhou, China, Nov. 2008.}

\newcommand{\ThankTwo}{This work was supported under Australian Research
Council's Discovery Projects funding scheme (project no.
DP0773898).}

\newcommand{\ThankThree}{Authors are with the College of Engineering and Computer Science, the
Australian National University, Canberra, ACT 0200, Australia.
Emails: \{xiangyun.zhou, parastoo.sadeghi, tharaka.lamahewa,
salman.durrani\}@anu.edu.au}

\title{Design Guidelines for Training-based MIMO Systems with Feedback}
\author{
\authorblockN{\textit{\AuthorOne,\:\AuthorTwo,\:\AuthorThree\:and\:\AuthorFour}
\thanks{\ThankOne}
\thanks{\ThankTwo}
\thanks{\ThankThree}
}}

\begin{document}

\pagestyle{empty}

\maketitle \vspace{-10mm}

\begin{abstract}

In this paper, we study the optimal training and data transmission
strategies for block fading multiple-input multiple-output (MIMO)
systems with feedback. We consider both the channel gain feedback
(CGF) system and the channel covariance feedback (CCF) system. Using
an accurate capacity lower bound as a figure of merit, we
investigate the optimization problems on the temporal power
allocation to training and data transmission as well as the training
length. For CGF systems without feedback delay, we prove that the
optimal solutions coincide with those for non-feedback systems.
Moreover, we show that these solutions stay nearly optimal even in
the presence of feedback delay. This finding is important for
practical MIMO training design. For CCF systems, the optimal
training length can be less than the number of transmit antennas,
which is verified through numerical analysis. Taking this fact into
account, we propose a simple yet near optimal transmission strategy
for CCF systems, and derive the optimal temporal power allocation
over pilot and data transmission.

\end{abstract}

\section{Introduction} \label{sec:intro}

\subsection{Background and Motivation}

The study of multiple-input multiple-output (MIMO) communication
systems can be broadly categorized based on the availability and
accuracy of channel state information (CSI) at the receiver or the
transmitter sides. Under the perfect CSI assumption at the receiver,
the MIMO channel information capacity and data transmission
strategies often have elegantly simple forms and many classical
results exist in the literature~\cite{foschini_gans_98,telatar_99}.
From~\cite{telatar_99,cover_06,visotsky_madhow_01,jafar_goldsmith_04,boche_jorswieck_04d,roh_rao_04,li_zhang_08}
we know that the MIMO information capacity with perfect receiver CSI
can be further increased if some form of CSI is fed back to the
transmitter. The transmitter CSI can be in the form of causal
channel gain feedback (CGF) or channel covariance feedback (CCF).

In practical communication systems with coherent detection, however,
the state of the MIMO channel needs to be estimated at the receiver
and hence, the receiver CSI is never perfect due to noise and time
variations in the fading channel. Taking the channel estimation
error into account, a widely-used capacity lower bound was
formulated in~\cite{hassibi_hochwald_03,yoo_goldsmith_06} for
independent and identically distributed (i.i.d.) MIMO channels, and
the optimal data transmission for CGF systems was studied
in~\cite{yoo_goldsmith_06}.

Pilot-symbol-assisted modulation (PSAM) has been used in many
practical communication systems, \eg in Global System for Mobile
Communications (GSM)~\cite{redl_95}. In PSAM schemes, pilot (or
training) symbols are inserted into data blocks periodically to
facilitate channel estimation at the receiver~\cite{cavers_91}. It
is noted that pilot symbols are not information-bearing signals.
Therefore, an important design aspect of communication systems is
the optimal allocation of resources (such as power and time) to
pilot symbols that results in the best tradeoff between the quality
of channel estimation and rate of information transfer. Three pilot
parameters under a system designer's control are: 1) spatial
structure of pilot symbols, 2) temporal power allocation to pilot
and data, and 3) the number of pilot symbols or simply training
length.

The optimal pilot design has been studied from an
information-theoretic viewpoint for non-feedback multi-antenna
systems of practical
interest~\cite{hassibi_hochwald_03,pohl_05,zhou_08}. For
non-feedback MIMO systems with i.i.d. channels, the authors
in~\cite{hassibi_hochwald_03} provided optimal solutions for all the
aforementioned design parameters by maximizing the derived capacity
lower bound. For CCF systems with correlated MIMO channels, the
optimal solution for the pilot's spatial structure was investigated
in~\cite{kotecha_sayeed_04,biguesh_gershman_06,pang_07}. However,
optimal solutions for the temporal pilot power allocation and
training length are generally unknown for MIMO systems with any form
of feedback. Some results were reported in~\cite{pang_08} for
rank-deficient channel covariance matrix known at the transmitter,
which are based on a relaxed capacity lower bound. However, this
relaxed capacity bound is generally loose for moderately to highly
correlated channels, which can render the provided solutions
suboptimal.

\subsection{Approach and Contributions}

In this paper, we are concerned with the optimal design of pilot
parameters for MIMO systems with various forms of feedback at the
transmitter. Our main design objectives are the optimal temporal
power allocation to pilot and data symbols, as well as the optimal
training length that maximize the rate of information transfer in
the channel. Our figure of merit is a lower bound on the ergodic
capacity of MIMO systems, which is an extension of those derived
in~\cite{yoo_goldsmith_06} from i.i.d channels to correlated
channels.

We address practical design questions such as: Are the simple
solutions provided in~\cite{hassibi_hochwald_03} for non-feedback
MIMO systems also optimal for systems with feedback? In CGF systems,
feedback delay is unavoidable. If the CGF takes $d$ symbol periods
to arrive at the transmitter, the transmitter can only utilize this
information after the first $d$ symbol periods. In this case, we
would like to know whether the optimal pilot design is significantly
affected by the feedback delay. Furthermore, for CCF systems with
correlated channels, the optimal training length may be shorter than
the number of transmit antennas, which is generally difficult to
solve analytically. In this case, we would like to know whether a
near-optimal, yet simple pilot and data transmission strategy
exists.

In this context, the main contributions of this paper are summarized
as follows.

\begin{itemize}

\item
For delayless CGF with i.i.d. channels, we show that the solutions
to the optimal temporal power allocation to pilot and data
transmission as well as the optimal training length coincide with
the solutions for non-feedback systems.

\item
For delayed CGF systems with i.i.d. channels, our numerical results
show that evenly distributing the power over the entire data
transmission (regardless of the delay time) gives near optimal
performance at practical signal-to-noise ratio (SNR). As a result,
the solutions to the optimal temporal power allocation to pilot and
data transmission, as well as the optimal training length for the
delayless system stay nearly optimal regardless of the delay time.

\item
For CCF systems with correlated channels, we propose a simple
transmission scheme, taking into account the fact that training
length $\Lp$ can be less than the number of transmit antennas. This
scheme only requires numerical optimization of $\Lp$ and does not
require numerical optimization over the spatial or temporal power
allocation over pilot and data transmission. Our numerical results
show that this scheme is very close to optimal. In addition, our
results show that optimizing $\Lp$ can result in a significant
capacity improvement for correlated channels.

\item
Using the proposed scheme for CCF systems, we find the solution to
the optimal temporal power allocation to pilot and data
transmission, which does not depend on the channel spatial
correlation under a mild condition on block length or SNR.
Therefore, the proposed transmission and power allocation schemes
for CCF systems give near optimal performance while having very low
computational complexity.

\end{itemize}

The rest of the paper is organized as follows. The PSAM transmission
scheme, channel estimation method, as well as an accurate capacity
lower bound for spatially correlated channels are presented in
Section~\ref{sec:SysMod}. The optimal transmission and power
allocation strategy for non-feedback systems are summarized in
Section~\ref{sec:nonfeedback}. The optimal transmission and power
allocation strategy for CGF and CCF systems are studied in
Section~\ref{sec:feedbackiid} and Section~\ref{sec:feedbackcor},
respectively. Finally, the main contributions of this paper are
summarized in Section~\ref{sec:summary}.

Throughout the paper, the following notations will be used: Boldface
upper and lower cases denote matrices and column vectors,
respectively. The matrix $\vI_N$ is the $N \times N$ identity
matrix. $\con{[\cdot]}$ denotes the complex conjugate operation, and
$\mct{[\cdot]}$ denotes the complex conjugate transpose operation.
The notation $E\{\cdot\}$ denotes the mathematical expectation.
$\text{tr}\{\cdot\}$, $|\cdot|$ and $\text{rank}\{\cdot\}$ denote
the matrix trace, determinant and rank, respectively.

\section{System Model} \label{sec:SysMod}

We consider a MIMO block-flat-fading channel model with input-output
relationship given by
\begin{eqnarray}\label{eq:MIMOSigMod}
        \vy =  \vH \vx + \vn,
\end{eqnarray}
where $\vy$ is the $\RxA \times 1$ received symbol vector, $\vx$ is
the $\TxA \times 1$ transmitted symbol vector, $\vH$ is the $\RxA
\times \TxA$ channel gain matrix, and $\vn$ is the $\RxA \times 1$
noise vector having zero-mean circularly symmetric complex Gaussian
(ZMCSCG) entries with variance $\sn$. Without loss in generality, we
let $\sn=1$. The entries of $\vH$ are also ZMCSCG with unit
variance. We consider spatial correlations among the transmit
antennas only. Therefore, $\vH = \vH_0 \vR^{1/2}_{\vH}$, where
$\vH_0$ has i.i.d. ZMCSCG entries with unit variance. The spatial
correlation at the transmitter is characterized by the covariance
matrix $\Rh = E\{\mct{\vH}\vH\}/\RxA$. In the case where the
channels are spatially independent, we have $\Rh = \vI_{\TxA}$. We
assume that $\Rh$ is a positive definite matrix and denote the
eigenvalues of $\Rh$ by $\vg = \mt{[g_1 \,\,g_2\,\,
\hdots\,\,g_{\TxA}]}$. Furthermore, we use the concept of
majorization to characterize the degree of channel spatial
correlation~\cite{chuah_02,boche_jorswieck_04c}, which is summarized
in Appendix~\ref{app:MeaCor}.

\subsection{Transmission Scheme}\label{sec:PSAM}

\begin{figure}[!t]
\centering\vspace{-0mm}
\includegraphics[width=0.7\columnwidth]{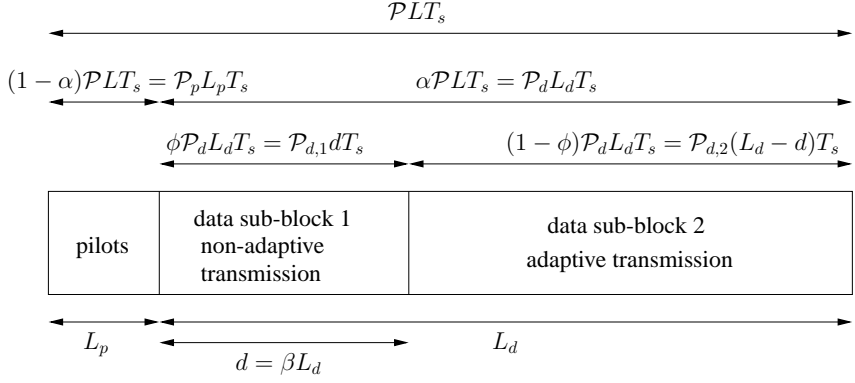}
\vspace{-6mm} \caption{An example of a transmission block of $L$
symbols in a system with delayed feedback. It consists of a training
sub-block, followed by two data sub-blocks. Temporal power
allocations are shown at the top and the length of each sub-block is
shown at the bottom.} \label{fig:block} \vspace{-3mm}
\end{figure}

Fig.~\ref{fig:block} shows an example of a transmission block of $L$
symbol periods in a PSAM scheme. The channel gains remain constant
over one block and change to independent realizations in the next
block. During each transmission block, each transmit antenna sends
$\Lp$ pilot symbols, followed by $\Ld$ ($=L-\Lp$) data symbols as
shown in Fig.~\ref{fig:block}. The receiver performs channel
estimation during the pilot transmission. For CGF systems, the
receiver feeds the channel estimates back to the transmitter once
per block to allow adaptive data transmission in the form of power
control. In practical scenarios, there is a time delay of $d$ symbol
periods before the transmitter receives the feedback information as
shown in Fig.~\ref{fig:block}. That is, the data transmission during
the first $d$ symbol periods is not adaptive to the channel, and
adaptive transmission is only available for the remaining $\Ld - d$
symbol periods. We define $\beta = d/\Ld$ as the feedback delay
factor. For CCF systems, less frequent feedback is required as the
channel correlation changes much slower than the channel gains.
Therefore, we do not consider feedback delay, \ie $d=0$. Note that
for non-feedback systems,  $d=\Ld$.

The total transmission energy per block is given by $\Pm L \Ts$ as
shown in Fig.~\ref{fig:block}, where $\Pm$ is the average power per
transmission and $\Ts$ is the symbol duration. We define the PSAM
power factor as the ratio of the total energy allocated to the data
transmission, denoted by $\alpha$. We also denote the power or SNR
per pilot and data transmission by $\Pp$ and $\Pd$\footnote{Ideally
for CGF systems, $\Pd$ should be larger for the transmission blocks
over which the channel is strong and smaller for blocks over which
the channel is weak. However, the results in~\cite{yoo_goldsmith_06}
suggest that this temporal data power adaptation provides little
capacity gain, hence it is not considered in this paper.},
respectively. Therefore, we have the following relationships.
\begin{eqnarray}\label{eq:power}
        \Pm L \Ts = \Pp \Lp \Ts + \Pd \Ld \Ts, \,\,\,
        \Pp = (1-\alpha) \frac{\Pm L}{\Lp}, \,\,\, \text{and} \,\,\, \Pd = \alpha \frac{\Pm L}{\Ld} .
\end{eqnarray}

For feedback systems with delay of $d$ symbol periods, the total
energy for data transmission $\Pd \Ld \Ts$ is further divided into
the non-adaptive data transmission sub-block and the adaptive data
transmission sub-block as shown in Fig.~\ref{fig:block}. We define
the data power division factor as the ratio of the total data energy
allocated to the non-adaptive sub-block, denoted by $\phi$.
Therefore, we have the following relationships.
\begin{eqnarray}\label{eq:DataPower}
        \Pd \Ld \Ts = \Pda d \,\Ts + \Pdb (\Ld-d) \Ts, \,\,\,
        \Pda = \frac{\phi}{\beta}\Pd, \,\,\, \text{and} \,\,\, \Pdb =
        \frac{1-\phi}{1-\beta}\Pd,
\end{eqnarray}
where $\Pda$ and $\Pdb$ are the power per transmission during the
non-adaptive and adaptive sub-blocks.

\subsection{Channel Estimation} \label{sec:ChaEst}

In each transmission block, the receiver performs channel estimation
during the pilot transmission. Combining the first $\Lp$ received
symbol vectors in a $\RxA~\times~\Lp$ matrix, we have
\begin{eqnarray}\label{eq:MIMOPilotBlockSigMod}
        \vY = \vH \vX_p + \vN,
\end{eqnarray}
where $\vX_p$ is the $\TxA \times \Lp$ pilot matrix and $\vN$ is the
$\RxA \times \Lp$ noise matrix.

Assuming the channel spatial correlation can be accurately measured
at the receiver, the channel gain $\vH$ can be estimated using the
linear minimum mean square error (LMMSE) estimator~\cite{kay_93}. We
denote the channel estimate and estimation error as $\hhat =
\hat{\vH}_0 \vR^{1/2}_{\hhat}$ and $\htilde = \tilde{\vH}_0
\vR^{1/2}_{\htilde}$ respectively, where $\hat{\vH}_0$ and
$\tilde{\vH}_0$ have i.i.d. ZMCSCG entries with unit variance.
$\hhat$ is given as~\cite{biguesh_gershman_06}
\begin{eqnarray}\label{eq:MIMOLMMSEestimator}
        \hhat = \vY(\mct{\vX_p}\Rh\vX_p+\vI_{\Lp})^{-1}\mct{\vX_p}\Rh.
\end{eqnarray}
The covariance matrix of the estimation error is given
by~\cite{biguesh_gershman_06}
\begin{eqnarray}\label{eq:MIMOLMMSECovariance}
        \Rhtilde = E\{\mct{\htilde}\htilde\}/\RxA=
        (\Rh^{-1}+\vX_p \mct{\vX_p})^{-1}.
\end{eqnarray}
From the orthogonality property of LMMSE estimator, we have
\begin{eqnarray}\label{eq:MIMOLMMSECovariance2}
        \Rhhat = E\{\mct{\hhat}\hhat\}/\RxA = \Rh - \Rhtilde.
\end{eqnarray}

\subsection{Ergodic Capacity Bounds} \label{sec:ErgCap}

The exact capacity expression under imperfect receiver CSI is still
unavailable. We consider a lower bound on the ergodic capacity for
systems using LMMSE channel
estimation~\cite{hassibi_hochwald_03,yoo_goldsmith_06}. In
particular, the authors in~\cite{yoo_goldsmith_06} derived a lower
bound and an upper bound for spatially i.i.d. channels. Here we
extend these results to spatially correlated channels as follows.

A lower bound on the ergodic capacity per channel use is given
by~\cite{yoo_goldsmith_06}
\begin{eqnarray}\label{eq:MIMOInsCapLB}
        \Clbl\!\!\!&=&\!\!\! E_{\hhat} \Big\{ \log_2 \Big|\vI_{\TxA} +\mct{\hhat}
        (\vI_{\RxA}+\vSig_{\htilde\vx})^{-1} \hhat \vQ \Big| \Big\},
\end{eqnarray}
where $\vQ = E\{\vx\mct{\vx}\}$ is the input covariance matrix, and
\begin{eqnarray}\label{eq:}
    \vSig_{\htilde\vx}\!\!\!&=&\!\!\!E\{\htilde\vx\mct{\vx}\mct{\htilde}\}\!
    =\! E \{\tilde{\vH}_0\vR^{1/2}_{\htilde} \vx\mct{\vx} \mct{(\vR^{1/2}_{\htilde})}
    \mct{\tilde{\vH}_0}\},\nonumber\\
    \!\!\!&=&\!\!\! E \Big\{ \,\text{tr}\{\vR^{1/2}_{\htilde} \vx\mct{\vx}
    \mct{(\vR^{1/2}_{\htilde})}\}\Big\}\vI_{\RxA}
    \!=\! {\text{tr}} \{\Rhtilde \vQ\}\vI_{\RxA},\nonumber
\end{eqnarray}
where we have used $E\{\tilde{\vH}_0 \vZ \mct{\tilde{\vH}_0}\} =
E\{\text{tr}\{\vZ\}\}\vI_{\RxA}$, given that $\tilde{\vH}_0$ has
i.i.d. entries with unit variance and is independent of $\vZ$.
Therefore, the ergodic capacity lower bound per channel use in
(\ref{eq:MIMOInsCapLB}) can be rewritten as
\begin{eqnarray}\label{eq:MIMOInsCapLB2}
       \Clbl \!\!\!&=&\!\!\! E_{\hhat} \Big\{\log_2 \Big|\vI_{\TxA} + (1+{\text{tr}} \{\Rhtilde \vQ\})^{-1}
        \mct{\hhat}\hhat \vQ \Big| \Big\}.
\end{eqnarray}

An upper bound on the ergodic capacity per channel use is given
by~\cite{yoo_goldsmith_06}
\begin{eqnarray}\label{eq:MIMOInsCapUB}
        \Cubl \!\!\!\!&=&\!\!\!\! E_{\hhat} \Big\{ \log_2 \Big| \pi e \vSig_{\vy|\hhat} \Big| \Big\}
        -E_{\vx} \Big\{ \log_2 \Big|\pi e (\vSig_{\htilde\vx|\vx}+\vI_{\RxA})\Big|
        \Big\},\nonumber
\end{eqnarray}
where
\begin{eqnarray}\label{eq:}
    \vSig_{\vy|\hhat} \!\!\!&=& \!\!\!E\{\vy\mct{\vy}|\hhat\}=\hhat\vQ\mct{\hhat}+{\text{tr}} \{\Rhtilde \vQ\}\vI_{\RxA}+\vI_{\RxA},\nonumber
\end{eqnarray}
and
\begin{eqnarray}\label{eq:}
    \vSig_{\htilde\vx|\vx}\!\!\!&=&\!\!\!E\{\htilde\vx\mct{\vx}\mct{\htilde}|\vx\}\!
    =\!E \{\tilde{\vH}_0\vR^{1/2}_{\htilde} \vx\mct{\vx}\mct{(\vR^{1/2}_{\htilde})} \mct{\tilde{\vH}_0}|\vx\},\nonumber\\
    \!\!\!&=&\!\!\!\text{tr}\{\vR^{1/2}_{\htilde} \vx\mct{\vx}\mct{(\vR^{1/2}_{\htilde})}\}\vI_{\RxA}
    \!=\!\mct{\vx}\Rhtilde\vx\vI_{\RxA}.\nonumber
\end{eqnarray}
Therefore, the ergodic capacity upper bound per channel use can be
written as
\begin{eqnarray}\label{eq:MIMOInsCapUB2}
        \Cubl\!\!\!&=&\!\!\!E_{\hhat} \Big\{ \log_2 \Big|\vI_{\TxA} + (1+{\text{tr}} \{\Rhtilde \vQ\})^{-1}
        \mct{\hhat}\hhat \vQ \Big| \Big\} +
        \RxA E_{\vx} \Big\{ \log_2 \frac{1+{\text{tr}} \{\Rhtilde \vQ\}}{1+\mct{\vx}\Rhtilde\vx} \Big\},\nonumber\\
        \!\!\!&=&\!\!\! \Clbl + \Cgap,
\end{eqnarray}
where $\Cgap$ is the difference between the upper bound and the
lower bound, which indicates the maximum error of the bounds. The
authors in~\cite{yoo_goldsmith_06} studied the tightness of the
bounds for i.i.d. channels. They observed that $\Cgap/\Clbl$ is
negligible for Gaussian inputs, hence the bounds are tight. We find
that this is also true for spatially correlated channels with LMMSE
estimation. Therefore, the capacity lower bound per channel use in
(\ref{eq:MIMOInsCapLB2}) is accurate enough to be used in our
analysis assuming Gaussian inputs. The average capacity lower bound
per transmission block is therefore given by
\begin{eqnarray}\label{eq:MIMOaveCapLB}
       \Clb = \frac{\Ld}{L}\Clbl = \frac{\Ld}{L} E_{\hhat} \Big\{\log_2 \Big|\vI_{\TxA} + (1+{\text{tr}} \{\Rhtilde \vQ\})^{-1}
        \mct{\hhat}\hhat \vQ \Big| \Big\}.
\end{eqnarray}
In this paper, the average capacity lower bound in
(\ref{eq:MIMOaveCapLB}) will be used as the figure of merit. We will
use ``capacity lower bound" and ``capacity" interchangeably
throughout the rest of this paper.

\section{Non-feedback Systems} \label{sec:nonfeedback}

\subsection{Spatially i.i.d. Channels} \label{sec:nonfeedbackiid}

The optimal pilot and data transmission scheme and optimal power
allocation for non-feedback systems with spatially i.i.d. channels
were studied in~\cite{telatar_99,hassibi_hochwald_03}, and their
main results are summarized as follows. The optimal transmission
strategy is to transmit orthogonal pilots and independent data among
the transmit antennas with spatially equal power allocation to each
antenna during both pilot and data transmission. The optimal PSAM
power factor $\alpha^*$ is given by
\begin{eqnarray}\label{eq:hassibi}
\alpha^* = \left\{ \begin{array}{ll}
 \gamma-\sqrt{\gamma(\gamma-1)}, &\mbox{ for $\Ld > \TxA$} \\
  \frac{1}{2}, &\mbox{ for $\Ld = \TxA$} \\
  \gamma+\sqrt{\gamma(\gamma-1)}, &\mbox{ for $\Ld < \TxA$}
       \end{array} \right.
\end{eqnarray}
where $\gamma = \frac{\TxA+ \Pm L}{\Pm L (1-\TxA/\Ld)}$. With the
optimal $\alpha$, the optimal training length is $\LpOpt = \TxA$.
For equal power allocation to pilot and data, \ie $\Pp=\Pd=\Pm$,
$\LpOpt$ should be found numerically.

\subsection{Spatially Correlated Channels} \label{sec:nonfeedbackcor}

In non-feedback systems where the transmitter does not know the
channel correlation, it is difficult to find the optimal resource
allocation and transmission strategies. Consequently, no results
have been found on the optimal or suboptimal solution to $\alpha^*$
and $\LpOpt$. Intuitively, the amount of training resource required
should reduce as the channels becomes more spatially correlated.
Therefore, one may use the solution to $\alpha^*$ and $\LpOpt$ for
i.i.d. channels as a robust strategy for correlated channels in
non-feedback systems. Similarly, one may still use the optimal
transmission strategies for i.i.d. channels to ensure a robust
system performance for correlated channels, which can be justified
by the following two theorems.

\begin{Theorem}\label{Theorem:0a}
{\em{For non-feedback systems with spatially correlated channels in
PSAM schemes, the transmission of orthogonal training sequences
among the transmit antennas with spatially equal power allocation
minimizes the channel estimation errors for the least-favourable
channel correlation, \ie using $\vX_p \mct{\vX_p} =
\frac{\Pp\Lp}{\TxA}\vI_{\TxA}$ is a robust training scheme.}}
\end{Theorem}

{\em{Proof:}} see Appendix~\ref{app:A}.

\begin{Theorem}\label{Theorem:0b}
{\em{For non-feedback systems with spatially correlated channels in
PSAM schemes, the transmission of i.i.d. data sequences among the
transmit antennas with spatially equal power allocation, \ie $\vQ =
\frac{\Pd}{\TxA}\vI_{\TxA}$, (a) maximizes the capacity for the
least-favourable channel correlation at sufficiently low SNR, and
(b) is the optimal transmission scheme at sufficiently high SNR.}}
\end{Theorem}

{\em{Proof:}} see~\cite{zhou_08b}.

{\em{Remark:}} From {\em{Theorem~\ref{Theorem:0a}}} and
{\em{Theorem~\ref{Theorem:0b}}}, we see that the optimal
transmission strategy for i.i.d. channels is also a robust choice
for correlated channels in non-feedback systems.

\section{Channel Gain Feedback (CGF) Systems}
\label{sec:feedbackiid}

In this section, we consider systems having a noiseless feedback
link from the receiver to the transmitter (\eg a low rate feedback
channel). After the receiver performs pilot-assisted channel
estimation, it feeds the channel estimates back to the transmitter.
Once the transmitter receives the estimated channel gains, it
performs spatial power adaptation accordingly. We consider the
channels to be spatially i.i.d.\footnote{We will provide some
discussion for CGF system with correlated channels in
Section~\ref{sec:cfgccf}.}. Since the data transmission utilizes all
the channels with equal probability, it is reasonable to have at
least as many measurements as the number of channels for channel
estimation, which implies that $\Lp \geq \TxA$.
From~\cite{hassibi_hochwald_03}, we know that the optimal training
consists of orthogonal pilots with equal power allocated to each
antenna.

\subsection{CGF System with No Feedback Delay} \label{sec:delaylessCGF}

Firstly, we study an ideal scenario in which the transmitter
receives the estimated channel gains at the start of the data
transmission, \ie $d=0$. For given $\Pd$, the ergodic capacity lower
bound per channel use in (\ref{eq:MIMOInsCapLB2}) can be rewritten
as
\begin{eqnarray}\label{eq:MIMOInsCapLB3}
       \Clbl &=& E_{\hat{\vH}_0} \Big\{\log_2 \Big|\vI_{\TxA} +
       \frac{\shhat}{1+\shtilde\Pd}\mct{\hat{\vH}_0}\hat{\vH}_0 \vQ \Big|
       \Big\},\nonumber\\
       &=& E_{\vlam} \Big\{ \sum_{i=1}^{\TxA} \log_2
       \Big(1+\frac{\shhat}{1+\shtilde\Pd}\lambda_i q_i\Big)\Big\},
\end{eqnarray}
where $\shtilde = \Big(1+\frac{\Pp \Lp}{\TxA}\Big)^{-1}$, $\shhat =
1 - \shtilde$, and $\vlam = \mt{[\lambda_1 \,\,\lambda_2\,\,
\hdots\,\,\lambda_{\TxA}]}$ denote the eigenvalues of
$\mct{\hat{\vH}_0}\hat{\vH}_0$. It was shown
in~\cite{yoo_goldsmith_06} that the capacity is maximized when the
matrix $\vQ$ has the same eigenvectors as
$\mct{\hat{\vH}_0}\hat{\vH}_0$. The eigenvalues of $\vQ$ can be
found via the standard water-filling given by
\begin{eqnarray}\label{eq:waterfillingIID}
       q_i = \Big[\eta -
       \Big(\frac{\shhat}{1+\shtilde\Pd}\lambda_i\Big)^{-1}\Big]^+
       \qquad {\text{with}} \qquad \sum_{i=1}^{\TxA} q_i = \Pd,
\end{eqnarray}
where $\eta$ represents the water level, and $[z]^+ \triangleq
\max\{z,0\}$. We refer to the number of non-zero $q_i$ as the number
of active eigen-channels, denoting this number by $m$. Therefore,
(\ref{eq:MIMOInsCapLB3}) can be reduced to
\begin{eqnarray}
       \Clbl &=& E_{\vlam} \Big\{ \sum_{i=1}^{m} \log_2
       \Big(\frac{\shhat}{1+\shtilde\Pd}\lambda_i\eta\Big)\Big\},\label{eq:MIMOInsCapLB4a}\\
            &=& E_{\vlam} \Big\{ \sum_{i=1}^{m} \log_2
       \Big(\frac{\shhat\Pd}{1+\shtilde\Pd} + \sum_{i=1}^{m}\lambda^{-1}_i \Big)
       + \sum_{i=1}^{m} \log_2 \frac{\lambda_i}{m}\Big\},\label{eq:MIMOInsCapLB4b}
\end{eqnarray}
where (\ref{eq:MIMOInsCapLB4b}) is obtained by substituting $\eta$
from (\ref{eq:waterfillingIID}) into (\ref{eq:MIMOInsCapLB4a}). It
should be noted that $E_{\vlam}$ in (\ref{eq:MIMOInsCapLB4a}) and
(\ref{eq:MIMOInsCapLB4b}) is the expectation over the $m$ largest
values in $\vlam$.

Using (\ref{eq:MIMOInsCapLB4b}), we now look for optimal value of
$\Pd$. The following two theorems summarize the results on the
optimal PSAM power factor $\alpha^*$ as well as the optimal training
length $\LpOpt$.

\begin{Theorem}\label{Theorem:1}
{\em{For delayless CGF systems with i.i.d. channels in PSAM schemes,
the optimal PSAM power factor $\alpha^*$ is given by
(\ref{eq:hassibi}).}}
\end{Theorem}

{\em{Proof:}} see Appendix~\ref{app:B1}.

\begin{Theorem}\label{Theorem:2}
{\em{For delayless CGF systems with i.i.d. channels in PSAM schemes
adopting the optimal PSAM power factor $\alpha^*$, the optimal
training length equals the number of transmit antennas, that is
$\LpOpt = \TxA$.}}
\end{Theorem}

{\em{Proof:}} see Appendix~\ref{app:B2}.

{\em{Remark:}} {\em{Theorem~\ref{Theorem:1}}} and
{\em{Theorem~\ref{Theorem:2}}} show that the optimal pilot design
for delayless CGF systems coincide with that for non-feedback
systems in Section~\ref{sec:nonfeedbackiid}. That is to say, one can
use the same design to achieve optimal performance in both
non-feedback and CGF systems.

\subsection{CGF System with Feedback Delay} \label{sec:delaylessCGF}

For practical systems, a finite duration of $d$ symbol periods is
required before feedback comes into effect at the transmitter as
shown in Fig.~\ref{fig:block}. Therefore, the transmitter has no
knowledge about the channel during the first data sub-block of $d$
transmissions, which is equivalent to non-feedback systems.
From~\cite{telatar_99}, we know that the transmitter should allocate
equal power to each transmit antenna during the first data sub-block
(or the non-adaptive sub-block). After receiving the estimated
channel gains, the transmitter performs spatial power water-filling
similar to Section~\ref{sec:delaylessCGF} during the second data
sub-block (or the adaptive sub-block) of length $\Ld-d$. Note that a
CGF system with $d=\Ld$ is equivalent to a non-feedback system.

In order to optimize PSAM power factor $\alpha$, we apply a
two-stage optimization approach. Firstly, we optimize the data power
division factor $\phi$ for a given total data power constraint.
Then, we optimize the PSAM power factor $\alpha$.

In general, we find that there is no closed-form solution for the
optimal data power division factor $\phi^*$. Furthermore when the
channel estimation error is large, the capacity lower bound is not
globally concave on $\phi \in [0,1]$. Nevertheless, the block length
$L$ of CGF systems is usually large (which will be discussed further
at the end of Section~\ref{sec:feedbackiid}). From the results on
the optimal PSAM power factor $\alpha^*$ and optimal training length
$\LpOpt$ in Section~\ref{sec:nonfeedbackiid} and
Section~\ref{sec:delaylessCGF}, we also expect that $\Pp \gg \Pm$
when $L \gg 1$. This implies that the channel estimation errors in
CGF systems are often small. Therefore, we can investigate the
optimal data power division assuming perfect channel estimation to
obtain some insights into the optimal solution for imperfect channel
estimation. In the following, we will see that a good approximation
of the optimal solution is given by $\phi^* \approx \beta$ for
practical SNR values under perfect channel estimation.

From (\ref{eq:DataPower}) we see that less power per transmission is
allocated to the non-adaptive sub-block (\ie $\Pda < \Pdb$) if $\phi
< \beta$, and {\textit{vice versa}}. The average capacity lower
bound for data transmission with perfect channel knowledge (\ie no
training) is given by
\begin{eqnarray}\label{eq:MIMOInsCapLB5}
       \Clbl &=& E_{\vlam} \Big\{ \beta \sum_{i=1}^{\TxA} \log_2
       \Big(1+\lambda_i \frac{\phi \Pd}{\beta \TxA}\Big)
       + (1-\beta) \sum_{i=1}^{\TxA} \log_2
       (1+\lambda_i q_i )\Big\},
\end{eqnarray}
where the water-filling solution for $q_i$ with water level $\nu$ is
given by
\begin{eqnarray}\label{eq:waterfillingIID2}
       q_i = [\nu - \lambda^{-1}_i]^+
       \qquad {\text{with}} \qquad \sum_{i=1}^{\TxA} q_i =
       \frac{1-\phi}{1-\beta}\Pd.
\end{eqnarray}

It can be shown that $\Clbl$ in (\ref{eq:MIMOInsCapLB5}) is concave
on $\phi \in [0,1]$.\footnote{This can be shown from the first and
second derivative of $\Clbl$ \wrt $\phi$ for any fixed number of
active eigen-channels $m$. In particular, one can show that
$\frac{\mathrm{d}\nu}{\mathrm{d}\phi}$ is continuous on $\phi \in
[0,1]$ and $\frac{\mathrm{d}^2\Clbl}{\mathrm{d}\phi^2} < 0$ for any
fixed $m$. Combining these two facts, one can conclude that $\Clbl$
is concave on $\phi \in [0,1]$. The detailed derivation is omitted
for brevity.} Using the Karush-Kuhn-Tucker (KKT)
conditions~\cite{boyd_04}, the optimal data power division factor
$\phi^*$ can be found as
\begin{eqnarray}\label{eq:OptPhiSol}
\left\{ \begin{array}{ll}
\phi^* = 0, &\mbox{ if $E_{\vlam}\{\lambda_i \} \leq E_{\vlam}\{\nu^{-1}\}$}\\
 \arg_{\phi} \, E_{\vlam} \Big\{ \beta \sum_{i=1}^{\TxA}
 \frac{\lambda_i}{\beta\TxA + \phi \lambda_i \Pd} - \nu^{-1} \Big\} = 0, &\mbox{ if
$E_{\vlam}\{ \lambda_i \} > E_{\vlam} \{\nu^{-1}\}$}
       \end{array} \right.
\end{eqnarray}
Note that the entries in $\vlam$ are the eigenvalues of a Wishart
matrix with parameter ($\TxA$, $\RxA$)~\cite{telatar_99}.

\begin{figure}[!t]
\centering\vspace{-3mm}
\includegraphics[width=0.8\columnwidth]{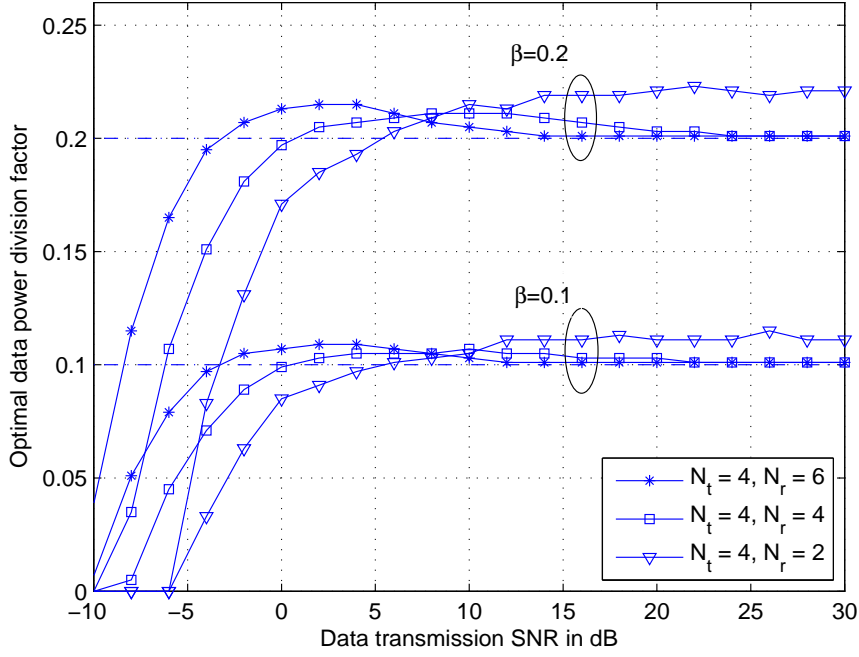}
\vspace{-6mm} \caption{The optimal data power division factor
$\phi^*$ vs. data transmission SNR $\Pd$ for different values of the
delay factor $\beta$ and antenna sizes. Perfect channel estimation
is assumed.} \label{fig:OptDataPowerSplit} \vspace{-3mm}
\end{figure}

Fig.~\ref{fig:OptDataPowerSplit} shows the optimal data power
division factor $\phi^*$ given by (\ref{eq:OptPhiSol}) versus data
transmission SNR $\Pd$ for different delay factors $\beta$ and
antenna sizes assuming perfect channel estimation. It can be seen
that $\phi^*$ quickly increases from 0 to $\beta$ at very low SNR.
For moderate to high SNR, $\phi^*$ stays above $\beta$ and converges
to $\beta$ as $\Pd \rightarrow \infty$.\footnote{$\phi^*$ for the
($\TxA = 4, \RxA = 2$) system starts to converge back to $\beta$ at
a higher SNR, which is not shown in
Fig.~\ref{fig:OptDataPowerSplit}. This is because the use of spatial
water-filling in data transmission gives a significant improvement
in the capacity when $\TxA > \RxA$.} More importantly, we see that
$\phi^*$ is close to $\beta$ at practical SNR range, \eg $\Pd>0$ dB.
Therefore, we conclude that $\phi = \beta$ is a near optimal
solution. From (\ref{eq:DataPower}) we see that $\phi = \beta$ is
actually the simplest solution which allocates the same amount of
power during each data transmission in both non-adaptive and
adaptive sub-blocks, \ie $\Pda = \Pdb = \Pd$. Furthermore, this
simple solution does not require the knowledge of the feedback delay
time.

Having $\phi^* \approx \beta$ for perfect channel estimation, we
argue that $\phi^* \approx \beta$ still holds for imperfect channel
estimation and will verify its optimality using numerical results.
This choice of $\phi$ leads to a simple solution for the optimal
PSAM power factor $\alpha^*$, as well as the optimal training length
$\LpOpt$ for delayed CGF system summarized in
{\em{Corollary~\ref{Corollary:1}}}, which can be shown by combining
the results in {\em{Theorem~\ref{Theorem:1}}},
{\em{Theorem~\ref{Theorem:2}}} and those for the non-feedback
systems summarized in Section~\ref{sec:nonfeedbackiid}.

\begin{Corollary}\label{Corollary:1}
{\em{For delayed CGF systems with i.i.d. channels in PSAM schemes,
temporally distributing equal power per transmission over both the
non-adaptive and adaptive data sub-blocks is a simple and efficient
strategy, \ie $\phi = \beta$. With this strategy, the optimal PSAM
power factor $\alpha^*$ and the optimal training length $\LpOpt$
coincide with those in the delayless case given in
Theorem~\ref{Theorem:1} and~\ref{Theorem:2}.}}
\end{Corollary}

\subsection{Numerical Results} \label{sec:feedbackiidresult}

Now, we present numerical results to illustrate the capacity gain
from optimizing the PSAM power factor. The numerical results also
validate the optimality of the transmission strategy in
{\em{Corollary~\ref{Corollary:1}}}.

\begin{figure}[!t]
\centering\vspace{-3mm}
\includegraphics[width=0.8\columnwidth]{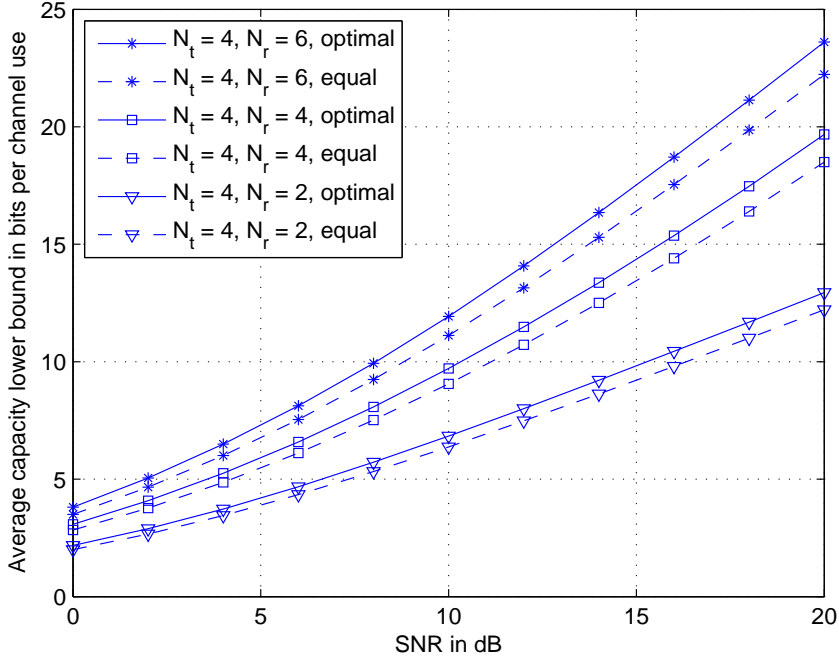}
\vspace{-6mm} \caption{Average capacity lower bound $\Clb$ in
(\ref{eq:MIMOaveCapLB}) vs. SNR $\Pm$ for delayless CGF systems
($\beta=0$) with i.i.d. channels and different antenna sizes. The
block length is $L=100$. Both optimal temporal power allocation to
pilot and data as well as equal power allocation are shown for
comparison. For optimal temporal power allocation, the training
length is $\LpOpt=4$; while for equal power allocation, the pilot
length is optimized numerically.}
\label{fig:capVsSNRnoDelayOptAndEquPilotPower} \vspace{-3mm}
\end{figure}

Fig.~\ref{fig:capVsSNRnoDelayOptAndEquPilotPower} shows the average
capacity lower bound $\Clb$ in (\ref{eq:MIMOaveCapLB}) versus SNR
$\Pm$ for delayless CGF systems (\ie $d=0$) with i.i.d. channels and
different antenna sizes. The solid lines indicate systems using
$\alpha^*$ and $\LpOpt$ ($\LpOpt=4$ in this case). The dashed lines
indicate systems using equal temporal power allocation and $\LpOpt$
found numerically. Comparing the solid and dashed lines, we see that
the capacity gain from optimal temporal power allocation is
approximately 9\% at 0 dB and 6\% at 20 dB for all three systems.
This range of capacity gain (5\% to 10\%) was also observed
in~\cite{hassibi_hochwald_03} for non-feedback systems which can be
viewed as an extreme case of delayed CGF system with $d=\Ld$. From
the results for the extreme cases, \ie $d=0$ and $d=\Ld$, we
conclude that the capacity gain from optimizing the PSAM power
factor is around 5\% to 10\% at practical SNR for delayed CGF
systems with i.i.d. channels.

\begin{figure}[!t]
\centering\vspace{-3mm}
\includegraphics[width=0.8\columnwidth]{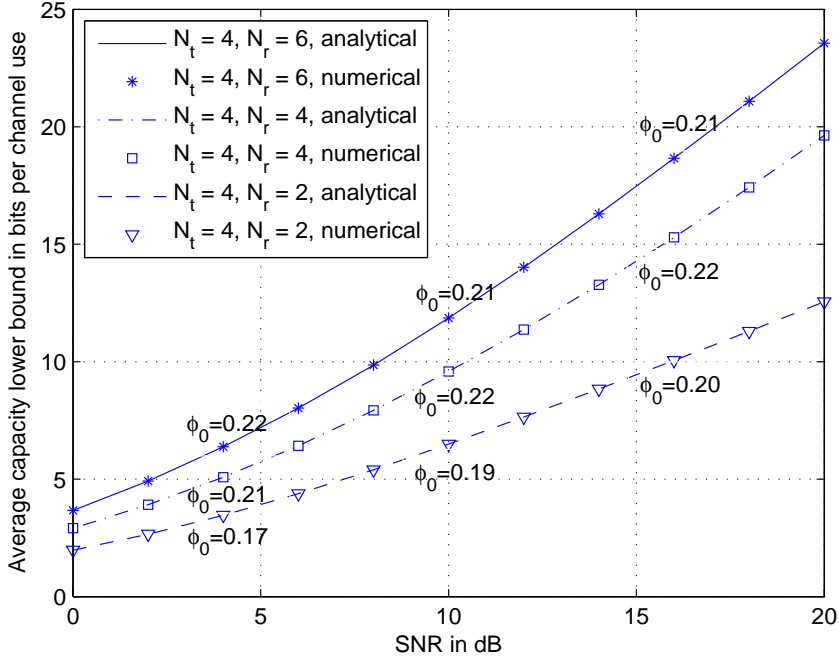}
\vspace{-6mm} \caption{Average capacity lower bound in
(\ref{eq:MIMOaveCapLB}) vs. SNR $\Pm$ for delayed CGF systems with
i.i.d. channels and different antenna sizes. Within a block length
of $L=100$, the training length is $\Lp=4$, followed by a
non-adaptive data transmission sub-block of length $d=20$ and an
adaptive data transmission sub-block of length $76$. The lines
indicate the use of $\phi=\beta=0.208$, and the markers indicate
optimal data power division factor found numerically.}
\label{fig:capVsSNRWithDelayAnalyticAndNumericOptDataPowerRatio}
\vspace{-3mm}
\end{figure}

We now consider delayed CGF systems to verify
{\em{Corollary~\ref{Corollary:1}}}.
Fig.~\ref{fig:capVsSNRWithDelayAnalyticAndNumericOptDataPowerRatio}
shows the average capacity lower bound $\Clb$ in
(\ref{eq:MIMOaveCapLB}) versus SNR $\Pm$ for delayed CGF systems
with i.i.d. channels and different antenna sizes. In this example, a
transmission block of length $L=100$ consists of a training
sub-block of $\Lp=4$ symbol periods, followed by a non-adaptive data
sub-block of $d=20$ symbol periods\footnote{The delay length $d$
takes into account the channel estimation and other processing time
at the receiver and transmitter, as well as the time spent on the
transmission of low-rate feedback.} and an adaptive data sub-block
of $\Ld-d=76$ symbol periods. Therefore, the delay factor $\beta =
0.208$. The lines indicate the use of $\phi=\beta$, and the markers
indicate optimal data power division found through numerical
optimization using $\Clb$ in (\ref{eq:MIMOaveCapLB}). The values of
$\phi^*$ for SNR = 4 dB, 10 dB and 16 dB are shown in the figure as
well. We see that the capacity difference between the system using
$\phi=\beta$ and $\phi=\phi^*$ is negligible. That is to say the use
of temporal equal power transmission over the entire data block is
near optimal for systems with channel estimation errors. We have
also confirmed that this trend is valid for a wide range of block
lengths (results are omitted for brevity). These results validate
{\em{Corollary~\ref{Corollary:1}}}.

It is noted that we have assumed the feedback link to be noiseless.
When noise is present, capacity that can be achieved by adaptive
transmission reduces as the noise in the feedback link increases.
The capacity reduction due to corrupted channel gain estimates was
studied in~\cite{baccarelli_05}. It was shown that the capacity
reduction can increase quickly with the noise in the estimated
channel gains. Therefore, a reliable feedback scheme which minimizes
the noise in the estimated channel gains is important for CGF
systems. Furthermore, CGF systems need frequent feedback
particularly when the block length is relatively small. This
requires a significant amount of feedback overhead in the reverse
link (from the receiver to the transmitter), which may cause a
direct reduction in the overall information rate, especially when
both the forward and the reverse links are operating at the same
time, \eg in cellular systems. Therefore, the CGF scheme may not be
appropriate in fast fading environments where the block length is
small.

\section{Channel Covariance Feedback (CCF) Systems}
\label{sec:feedbackcor}

As discussed in the previous subsection, CGF systems require
frequent use of feedback due to the rapid change in the channel
gains. On the other hand, the statistics of the channel gains change
much slower than the channel gains themselves. As a result, it is
practical for the receiver to accurately measure the channel
covariance matrix and feed it back to the transmitter at a much
lower frequency with negligible feedback overhead and delay. Note
that for completely i.i.d. channels, there is no need for CCF. In
this section, we consider CCF systems with spatially correlated
channels and investigate the optimal pilot and data transmission
strategy, as well as the optimal power allocation.

\subsection{Proposed Transmission Scheme} \label{sec:proposedtxscheme}

Intuitively, the amount of training resource required for spatially
correlated channels should be less than that for i.i.d. channels, as
spatial correlation reduces the uncertainty in the channel gains.
From~\cite{hassibi_hochwald_03}, we know for i.i.d. channels that
the optimal training length $\LpOpt$ equals the number of transmit
antennas provided that the optimal PSAM power factor $\alpha^*$ is
used. Therefore, we expect that $\LpOpt \leq \TxA$ for correlated
channels if we optimize $\alpha$. However, most studies on the
optimal pilot design for correlated channels assume $\Lp \geq
\TxA$~\cite{kotecha_sayeed_04,biguesh_gershman_06,pang_07}. It was
shown in~\cite{biguesh_gershman_06} that the optimal training
strategy is to train along the eigenvectors of the channel
covariance matrix with training power being waterfilled according to
the eigenvalues of the channel covariance matrix. Since we expect
$\Lp \leq \TxA$, we modify the training strategy such that only the
$\Lp$ strongest eigen-channels are trained.

We perform eigenvalue decomposition on $\Rh$ as $\Rh = \vU \vG
\mct{\vU}$, and let the eigenvalues of $\Rh$ be sorted in descending
order in $\vg = \mt{[g_1 \,\,g_2\,\, \hdots\,\,g_{\TxA}]}$. The
optimal training sequence which minimizes the channel estimation
errors (\ie $\text{tr}\{\Rhtilde\}$) has the property that the
eigenvalue decomposition of $\vX_p \mct{\vX_p}$ is given by $\vX_p
\mct{\vX_p} = \vU \vP \mct{\vU}$~\cite{biguesh_gershman_06}, where
$\vP$ is a diagonal matrix. The entries of $\vP$ which minimize the
channel estimation errors follow a water-filling solution given by
\begin{eqnarray}\label{eq:waterfillingtrain}
p_i = \left\{ \begin{array}{ll}
 [\mu - g^{-1}_i]^+, \,\,\,\,\,\, i=1,\hdots,\Lp, &\mbox{ with $\sum_{i=1}^{\Lp} p_i = \Pp\Lp$} \\
 0, \,\,\,\,\,\,\,\,\,\,\,\,\,\, i=\Lp+1,\hdots,\TxA,
       \end{array} \right.
\end{eqnarray}
where $\mu$ is the water level and $\vp = \mt{[p_1 \,\,p_2\,\,
\hdots\,\,p_{\TxA}]}$ are the eigenvalues of $\vX_p \mct{\vX_p}$. In
practice, the transmitter can ensure that the number of non-zero
$p_i$ equals $\Lp$ by changing $\Lp$ accordingly.

For data transmission, it was shown that the optimal strategy is to
transmit along the eigenvectors of $\Rh$ under the perfect channel
estimation~\cite{visotsky_madhow_01,jafar_goldsmith_04,boche_jorswieck_04d}.
With channel estimation errors, one strategy is to transmit data
along the eigenvectors of $\Rhhat$. With the proposed training
sequence, it is easy to show from (\ref{eq:MIMOLMMSECovariance}) and
(\ref{eq:MIMOLMMSECovariance2}) that the eigenvectors of $\Rhtilde$
and $\Rhhat$ are the same as those of $\Rh$. Therefore, the
eigenvalue decomposition of $\Rhhat$ can be written as $\Rhhat = \vU
\hat{\vG} \mct{\vU}$, and we set $\vQ = \vU \hat{\vQ} \mct{\vU}$
where $\hat{\vQ}$ is a diagonal matrix with entries denoted by $q_i,
\,\forall \,i=1,...,\TxA$.

However, there is no closed-form solution to the optimal spatial
power allocation even with perfect channel
estimation~\cite{visotsky_madhow_01,jafar_goldsmith_04,boche_jorswieck_04d}.
Following the proposed training scheme, we propose to transmit data
through the $\Lp$ trained eigen-channels with equal power. That is
\begin{eqnarray}\label{eq:waterfillingdata}
q_i = \left\{ \begin{array}{ll}
 \Pd/\Lp, \,\,\,\,\,\, i=1,\hdots,\Lp, \\
  0, \,\,\,\,\,\,\,\, i=\Lp+1,\hdots,\TxA.
       \end{array} \right.
\end{eqnarray}

For the proposed training and data transmission scheme, the capacity
lower bound per channel use in (\ref{eq:MIMOInsCapLB2}) reduces to
\begin{eqnarray}\label{eq:MIMOInsCapLB6}
       \Clbl \!\!\!&=&\!\!\! E_{\hat{\vH}_0} \Big\{\log_2 \Big|\vI_{\TxA} +
       \mct{\hat{\vH}_0}\hat{\vH}_0 \hat{\vG} \hat{\vQ} (1+\mu^{-1}\Pd)^{-1}
       \Big| \Big\},
\end{eqnarray}
where the (diagonal) entries of $\hat{\vG}$ are given by $\hat{g}_i
= g_i - \mu^{-1}, \,\forall \,i=1,...,\Lp$ and $\hat{g}_i = 0,
\,\forall \,i=\Lp+1,..,\TxA$, which is derived from
(\ref{eq:MIMOLMMSECovariance}), (\ref{eq:MIMOLMMSECovariance2}) and
(\ref{eq:waterfillingtrain}).

\subsection{Optimal Temporal Power Allocation} \label{sec:OptTemPowAll}

Now, we investigate the optimal PSAM power factor $\alpha^*$ using
the capacity lower bound given in (\ref{eq:MIMOInsCapLB6}). The
result is summarized in the following theorem.

\begin{Theorem}\label{Theorem:3}
{\em{For CCF systems in PSAM schemes with the transmission strategy
proposed in Section~\ref{sec:proposedtxscheme}, the optimal PSAM
power factor $\alpha^*$ is given by (\ref{eq:hassibi}) with $\gamma
= \frac{\Ld}{\Ld - \Lp}$, provided that $\Pm
L\gg\sum_{i=1}^{\Lp}g^{-1}_i$.}}
\end{Theorem}

{\em{Proof:}} see Appendix~\ref{app:C}.

{\em{Remark:}} It is noted that $\gamma$ in the optimal solution in
{\em{Theorem~\ref{Theorem:3}}} is essentially the same as the one
given in Section~\ref{sec:nonfeedbackiid} when $\Pm L \gg 1$. The
condition of $\Pm L\gg\sum_{i=1}^{\Lp}g^{-1}_i$ can be easily
satisfied when the block length is not too small or the SNR is
moderate to high (\ie $\Pm L\gg1$), and the spatial correlation
between any trained channels is not close to 1. Therefore, the
result in {\em{Theorem~\ref{Theorem:3}}} applies to many practical
scenarios. It is important to note that the optimal PSAM power
factor $\alpha^*$ given in {\em{Theorem~\ref{Theorem:3}}} does not
depend on the channel spatial correlation, provided the condition is
met. In other words, this unique design is suitable for a relatively
wide range of channel spatial correlation.

\vspace{3mm} \fbox{\parbox[c][4.8cm][c]{15cm}{

The following steps describe the algorithm for transmission design
of CCF systems:

1. For each $\Lp$ ($\Lp \leq \TxA$), design the pilot and data
transmission according to Section~\ref{sec:proposedtxscheme}.

2. Perform temporal power allocation to pilot and data according to
Section~\ref{sec:OptTemPowAll}.

3. Numerically compare the capacity lower bound in
(\ref{eq:MIMOaveCapLB}) for different $\Lp$ and choose $\LpOpt$
which maximizes the capacity. }}

\vspace{3mm}

\subsection{A Special Case: Beamforming} \label{sec:beamforming}

Beamforming is a special case of the proposed transmission scheme
where only the strongest eigen-channel is used, \ie $\Lp=1$. The use
of beamforming significantly reduces the complexity of the system as
it allows the use of well-established scalar codec technology and
only requires the knowledge of the strongest eigen-channel (not the
complete channel statistics)~\cite{jafar_goldsmith_04}. For
beamforming transmission, the capacity lower bound in
(\ref{eq:MIMOInsCapLB6}) reduces to
\begin{eqnarray}\label{eq:MIMOInsCapLBbeamforming}
       \Clbl \!\!\!&=&\!\!\! E_{\hat{\vh}_0} \Big\{\log_2 \Big(1 +
       \mct{\hat{\vh}_0}\hat{\vh}_0 \frac{(g_{\text{max}}-\mu^{-1})\Pd}{1+\mu^{-1}\Pd}
       \Big) \Big\},\nonumber\\
        &=&\!\!\! E_{\hat{\vh}_0} \Big\{\log_2 \Big(1 +
       \mct{\hat{\vh}_0}\hat{\vh}_0 \frac{g_{\text{max}}\Pp\Pd}{g^{-1}_{\text{max}}+\Pp + \Pd}
       \Big) \Big\},
\end{eqnarray}
where $\hat{\vh}_0$ is a $\RxA \times 1$ vector with i.i.d. ZMCSCG
and unit variance entries, $g_{\text{max}}$ is the largest
eigenvalue in $\vg$, and $\mu = \Pp + g^{-1}_{\text{max}}$ which can
be found by letting $\Lp=1$ in (\ref{eq:waterfillingtrain}).

\begin{Theorem}\label{Theorem:5}
{\em{For CCF systems in PSAM schemes with beamforming, the optimal
PSAM power factor $\alpha^*$ is given in (\ref{eq:hassibi}) with
$\gamma = \frac{1+g_{\text{max}}\Pm L}{g_{\text{max}}\Pm L
(L-2)/(L-1)}$.}}
\end{Theorem}

{\em{Proof:}} The proof can be obtained by letting $\Lp=1$ and
$g_i=g_{\text{max}}$ in the proof of {\em{Theorem~\ref{Theorem:3}}}.
$\Box$

{\em{Remark:}} It can be shown for the beamforming case that
$\frac{\mathrm{d}\alpha^*}{\mathrm{d}g_{\text{max}}} > 0$.
Therefore, the optimal PSAM power factor $\alpha^*$ increases as the
channel spatial correlation increases, that is to say, more power
should be allocated to data transmission when the channels become
more correlated. When $\Pm L \gg 1$, $\gamma$ reduces to
$\frac{L-1}{L-2}$, hence $\alpha^*$ does not depend on the channel
correlation.

\subsection{Numerical Results} \label{sec:feedbackcorresult}

For numerical analysis, we choose the channel covariance matrix to
be in the form of $[\Rh]_{ij} = \rho^{|i-j|}$, where $\rho$ is
referred to as the spatial correlation
factor~\cite{yoo_yoon_goldsmith_04,biguesh_gershman_06}. Our
numerical results validate the solution to the optimal PSAM power
factor given in {\em{Theorem~\ref{Theorem:3}}} and
{\em{Theorem~\ref{Theorem:5}}}. The results also show that
optimizing the training length can significantly improve the
capacity, and the simple transmission scheme proposed in
Section~\ref{sec:proposedtxscheme} gives near optimal performance.

\begin{figure}[!t]
\centering\vspace{-3mm}
\includegraphics[width=0.8\columnwidth]{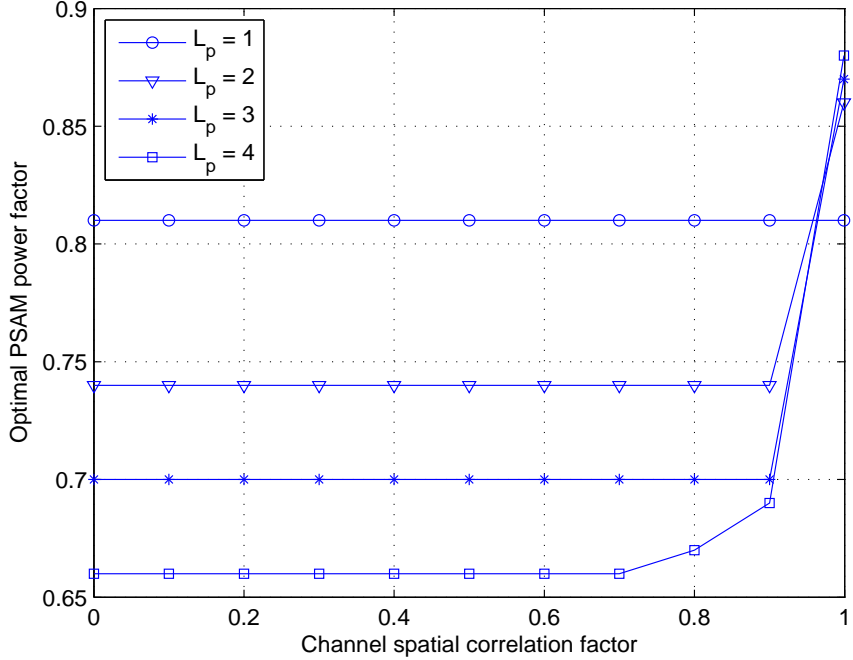}
\vspace{-6mm} \caption{Optimal PSAM power factor $\alpha^*$ vs.
channel spatial correlation factor $\rho$ for CCF $4 \times 4$
systems with a block length of $L=20$ and SNR = 10 dB. All values of
$\alpha^*$ are found numerically.}
\label{fig:optPowerRatioT20snr10dB} \vspace{-3mm}
\end{figure}

Fig.~\ref{fig:optPowerRatioT20snr10dB} shows the optimal PSAM power
factor $\alpha^*$ found numerically versus the channel correlation
factor $\rho$ for CCF $4 \times 4$ systems with a block length of
$L=20$ and SNR of 10 dB. We see that $\alpha^*$ remains constant
before the correlation factor gets close to 1 for $\Lp >1 $, and
this value of $\alpha^*$ is the same as the analytical value
computed from {\em{Theorem~\ref{Theorem:3}}}. For the beamforming
case where $\Lp=1$, we see that $\alpha^*$ does not depend on the
channel correlation, which agrees with our earlier observation from
{\em{Theorem~\ref{Theorem:5}}}. Similar to CGF systems, we have also
compared the capacity achieved using $\alpha^*$ and that using equal
power allocation over pilot and data, and the same trend is observed
(results are omitted for brevity), that is, capacity gain from
optimizing PSAM power factor is around 5\% to 10\% at practical SNR.

\begin{figure}[!t]
\centering\vspace{-3mm}
\includegraphics[width=0.8\columnwidth]{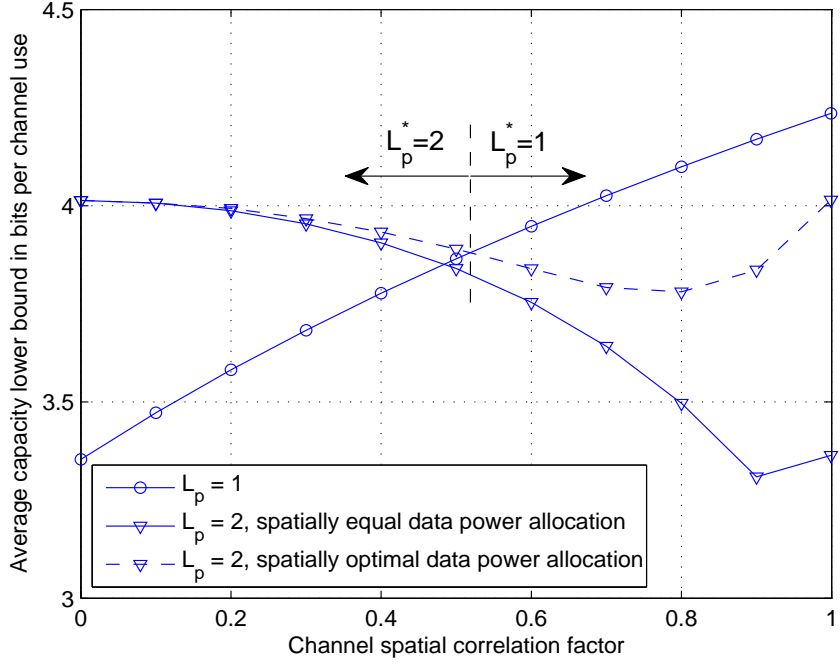}
\vspace{-6mm} \caption{Average capacity lower bound $\Clb$ in
(\ref{eq:MIMOaveCapLB}) vs. channel spatial correlation factor
$\rho$ for CCF $2 \times 2$ systems with a block length of $L=20$
and SNR of 10 dB. Training length of $\Lp=1$ and $\Lp=2$ are shown.
For $\Lp=2$, both spatial equal data power allocation (dashed lines)
and optimal data power allocation found numerically (solid lines)
are shown.} \label{fig:CapVsCorrSnr10dBcompOptPower} \vspace{-3mm}
\end{figure}

In our proposed transmission scheme for CCF systems, spatially equal
power allocation is used for data transmission. Here we illustrate
the optimality of this simple scheme in
Fig.~\ref{fig:CapVsCorrSnr10dBcompOptPower}, which shows the average
capacity lower bound $\Clb$ in (\ref{eq:MIMOaveCapLB}) versus
channel correlation factor $\rho$ for CCF $2 \times 2$ systems. We
compute the capacity achieved using $\Lp=1$, and $\Lp=2$ with
spatially equal power allocation for data transmission (solid line)
and optimal power allocation found numerically (dashed line) for a
block length of $L=20$.\footnote{We see that the capacity increases
with channel spatial correlation in the case of beamforming, while
it is not monotonic for $\Lp=2$. These observations were explained
in~\cite{zhou_08b} using Schur-convexity of capacity in the channel
correlation.} We also indicate the critical $\rho$ at which $\LpOpt$
changes from 2 to 1 in Fig.~\ref{fig:CapVsCorrSnr10dBcompOptPower}.
It is clear that the capacity loss from spatially optimal power
allocation to spatially equal power allocation increases as $\rho$
increases. At the critical $\rho$, this capacity loss is only around
1.5\%. We also studied the results for different values of block
lengths and the same trend was found (results are omitted for
brevity). These results imply that our proposed transmission scheme
is very close to optimal provided that the training length is
optimized.

\begin{figure}[!t]
\centering\vspace{-3mm}
\includegraphics[width=0.8\columnwidth]{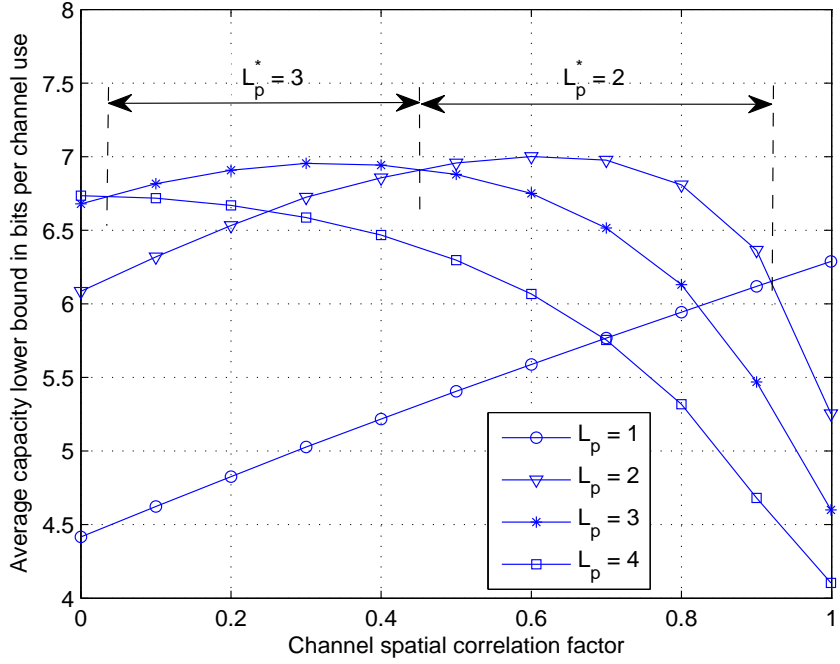}
\vspace{-6mm} \caption{Average capacity lower bound $\Clb$ in
(\ref{eq:MIMOaveCapLB}) vs. channel spatial correlation factor
$\rho$ for CCF $4 \times 4$ systems with a block length of $L=20$
and SNR = 10 dB. The optimal PSAM power factor $\alpha^*$ is used in
all results.} \label{fig:CapLBoptPowerT20snr10dB} \vspace{-3mm}
\end{figure}

Fig.~\ref{fig:CapLBoptPowerT20snr10dB} shows the average capacity
lower bound $\Clb$ in (\ref{eq:MIMOaveCapLB}) versus the channel
correlation factor $\rho$ for CCF $4 \times 4$ systems with a block
length of $L=20$ and SNR of 10 dB. The optimal PSAM power factor
$\alpha^*$ shown in Fig.~\ref{fig:optPowerRatioT20snr10dB} is used
in the capacity computation. Comparing the capacity with different
training lengths, we see that $\LpOpt$ decreases as the channel
becomes more correlated. More importantly, the capacity gain from
optimizing the training length according to the channel spatial
correlation can be significant. For example, the capacity at
$\rho=0.5$ using $\Lp=4$ (which is optimal for i.i.d. channels) is
approximately 6.3 bits per channel use, while the capacity at
$\rho=0.5$ using $\LpOpt=2$ is around 7 bits per channel use, that
is to say, optimizing training length results in a capacity
improvement of 11\% at $\rho=0.5$. Moreover, the capacity
improvement increases as channel correlation increases. The same
trends are found for different values of block lengths, although the
capacity improvement by optimizing the training length reduces as
the block length increases (results are omitted for brevity).
Therefore, it is important to numerically optimize the training
length for correlated channels at small to moderate block lengths.

Furthermore, one can record the range of $\rho$ for each value of
$\LpOpt$ from Fig.~\ref{fig:CapLBoptPowerT20snr10dB}, and observe
the value of $\alpha^*$ in the corresponding range of $\rho$ in
Fig.~\ref{fig:optPowerRatioT20snr10dB}. It can be seen that within
the range of $\rho$ where a given $\Lp$ is optimal, the value of
$\alpha^*$ for the given $\Lp$ is a constant given by
{\em{Theorem~\ref{Theorem:3}}} provided that $\Pm L\gg 1$. That is
to say, the condition in {\em{Theorem~\ref{Theorem:3}}} (\ie $\Pm
L\gg\sum_{i=1}^{\Lp}g^{-1}_i$) can be simplified to $\Pm L \gg 1$
provided that the training length is optimized.

\subsection{Hybrid CGF and CCF Systems} \label{sec:cfgccf}

After studying the optimal transmission and power allocation
strategy for CGF systems with i.i.d. channels and CCF system with
correlated channels, we provide some discussion on systems utilizing
both CGF and CCF with correlated channels. For spatially correlated
channels, the optimal training follows a water-filling solution
according to the channel covariance, and the optimal data
transmission follows a water-filling solution according to the
estimated channel gains. The two different water-filling solutions
make the problem of optimizing the PSAM power factor mathematically
intractable. Furthermore, the optimal training length $\LpOpt$ may
be smaller than the number of transmit antennas, and needs to be
found numerically. However, from the results for CGF systems with
i.i.d. channels in Section~\ref{sec:feedbackiid} and CCF system with
correlated channel in Section~\ref{sec:feedbackcor}, one may expect
that a good solution for the optimal PSAM power factor $\alpha^*$ in
the hybrid system is given in {\em{Theorem~\ref{Theorem:3}}}.

\section{Summary of Results}
\label{sec:summary}

In this paper, we have studied block fading MIMO systems with
feedback in PSAM transmission schemes. Two typical feedback systems
are considered, namely the channel gain feedback and the channel
covariance feedback systems. Using an accurate capacity lower bound
as the figure of merit, we have provided the solutions for the
optimal power allocation to training and data transmission as well
as the optimal training length. Table~\ref{table:guidelines}
summarizes the design guidelines for both non-feedback systems and
feedback systems.

\begin{table}[!ht]
\caption{Summary of Design Guidelines} \centering
\begin{tabular}{l|l|l|l} \hline
     System & Channel & Design Guidelines & Reference\\ \hline
     & \multirow{4}{*}{i.i.d.} & $\bullet\,$Transmit orthogonal pilots among antennas with spatially equal power. &  \multirow{4}{*}{\cite {telatar_99,hassibi_hochwald_03}}\\
    Non-& & $\bullet\,$Transmit independent data among antennas with spatially equal power. &\\
     feedback & & $\bullet\,$The optimal PSAM power factor $\alpha^*$ is given by (\ref{eq:hassibi}) with $\gamma = \frac{\TxA+ \Pm L}{\Pm L
     (1-\TxA/\Ld)}$. &\\
    & & $\bullet\,$The optimal training length $\LpOpt$ equals the number of transmit antennas $\TxA$.&\\ \cline{2-4}
     & correlated & $\bullet\,$Use the designs for i.i.d.
     channels as a robust choice. & Sec.~\ref{sec:nonfeedbackcor}\\ \hline
     \multirow{6}{*}{CGF} & \multirow{6}{*}{i.i.d.}  & $\bullet\,$Transmit orthogonal pilots with spatially equal power. & \multirow{6}{*}{Sec.~\ref{sec:feedbackiid}}\\
    & & $\bullet\,$Transmit independent data with spatially equal power in data sub-block 1 &\\
    & & $\,\,\,$and spatial power water-filling in data sub-block 2 (see Fig.~\ref{fig:block}).&\\
    & & $\bullet\,$Distribute equal power per transmission throughout data sub-blocks 1 and 2.&\\
     & & $\bullet\,$$\alpha^*$ and $\LpOpt$ for non-feedback system are (near) optimal for (delayed) CGF system. &\\ \hline
     \multirow{5}{*}{CCF} & \multirow{5}{*}{correlated} & $\bullet\,$For a given $\Lp$ ($\Lp \leq \TxA$), transmit pilots along the $\Lp$ strongest
     & \multirow{5}{*}{Sec.~\ref{sec:feedbackcor}}\\
     & & $\,\,\,$eigen-channels with spatial power water-filling according to (\ref{eq:waterfillingtrain}).& \\
     & & $\bullet\,$Transmit data along the $\Lp$ trained eigen-channels with spatially equal power. &\\
     & & $\bullet\,$$\alpha^*$ is given by (\ref{eq:hassibi}) with $\gamma = \frac{\Ld}{\Ld - \Lp}$, provided that $\Pm
    L\gg\sum_{i=1}^{\Lp}g^{-1}_i$. &\\
    & & $\bullet\,$$\LpOpt$ should be numerically optimized. &\\
    & & $\bullet\,$For beamforming (\ie $\Lp = 1$), $\alpha^*$ is given by (\ref{eq:hassibi}) with
    $\gamma = \frac{1+g_{\text{max}}\Pm L}{g_{\text{max}}\Pm L(L-2)/(L-1)}$. &\\ \hline
\end{tabular}
\label{table:guidelines}
\end{table}

\appendices

\section{A Measure of Channel Spatial Correlation}
\label{app:MeaCor}

A vector $\va = \mt{[a_1 \,\,a_2\,\, \hdots\,\,a_n]}$ is said to be
majorized by another vector $\vb = \mt{[b_1 \,\,b_2\,\,
\hdots\,\,b_n]}$ if
\begin{eqnarray}
        \sum_{i=1}^{k}a_i \!\leq\! \sum_{i=1}^{k}b_i, \,\,\,\,
        k=1,\hdots,n-1, \,\,\,\,\,\, \text{and} \,\,\,\,\sum_{i=1}^{n}a_i \!= \!\sum_{i=1}^{n}b_i,
\end{eqnarray}
where the elements in both vectors are sorted in descending
order~\cite{marshall_olkin_79}. We denote the relationship as $\va
\prec \vb$. Any real-valued function $\Phi$, defined on a vector
subspace, is said to be Schur-convex, if $\va \prec \vb$ implies
$\Phi(\va) \leq \Phi(\vb)$~\cite{marshall_olkin_79}. Similarly
$\Phi$ is Schur-concave, if $\va \prec \vb$ implies $\Phi(\va) \geq
\Phi(\vb)$. Following~\cite{boche_jorswieck_04c}, we have the
following definition:\vspace{3mm}

\begin{Definition}
Let $\va$ contain the eigenvalues of a channel covariance matrix
such as $\vR_{\va}$, and $\vb$ contain the eigenvalues of another
channel covariance matrix $\vR_{\vb}$. The elements in both vectors
are sorted in descending order. Then $\vR_{\va}$ is less correlated
than $\vR_{\vb}$ if and only if $\va \prec \vb$.
\end{Definition}

\section{Proof of Theorem 1}
\label{app:A}

This is a max-min problem where the MSE of the channel estimates is
to be minimized by $\vX_p \mct{\vX_p}$ and to be maximized by $\Rh$.
We need to show that $\text{inf}_{\vX_p \mct{\vX_p}}
\,\,\text{sup}_{\Rh}\, \text{tr}\{\Rhtilde\}$ is achieved by
orthogonal pilot sequence with equal power allocated among the
transmit antennas, \ie $\vX_p \mct{\vX_p} =
\frac{\Pp\Lp}{\TxA}\vI_{\TxA}$, assuming $\Lp \geq \TxA$.

From (\ref{eq:MIMOLMMSECovariance}) we see that
\begin{eqnarray}\label{eq:MSE1}
       \text{sup}_{\Rh}\, \text{tr}\{\Rhtilde\} &\geq& \text{tr}\{(\vI_{\TxA} + \vX_p
       \mct{\vX_p})^{-1}\}=\sum_{i=1}^{\TxA}(1+p_i)^{-1},
\end{eqnarray}
where $\vp = \mt{[p_1 \,\,p_2\,\, \hdots\,\,p_{\TxA}]}$ are the
eigenvalues of $\vX_p \mct{\vX_p}$. Since the sum of a convex
function of $p_i$ is Schur-convex in $\vp$~\cite{marshall_olkin_79},
we conclude that (\ref{eq:MSE1}) is Schur-convex in $\vp$. Since
$\text{tr}\{\vX_p \mct{\vX_p}\} = \Pp \Lp$, we have
\begin{eqnarray}\label{eq:MSE2}
       \text{sup}_{\Rh}\, \text{tr}\{\Rhtilde\} &\geq&
       \sum_{i=1}^{\TxA}\Big(1+\frac{\Pp\Lp}{\TxA}\Big)^{-1},
\end{eqnarray}
where we have used $\vX_p \mct{\vX_p} =
\frac{\Pp\Lp}{\TxA}\vI_{\TxA}$. Note that (\ref{eq:MSE2}) holds for
any $\vX_p \mct{\vX_p}$. On the other hand
\begin{eqnarray}\label{eq:MSE3}
       \text{inf}_{\vX_p \mct{\vX_p}}
\,\,\text{sup}_{\Rh}\, \text{tr}\{\Rhtilde\} &\leq&
\,\,\text{sup}_{\Rh}\, \text{tr}\Big\{\Big(\Rh^{-1} + \frac{\Pp\Lp}{\TxA}\vI_{\TxA}\Big)^{-1}\Big\},\nonumber\\
       &=&\,\,\text{sup}_{\Rh}\, \sum_{i=1}^{\TxA}\Big(g^{-1}_i + \frac{\Pp\Lp}{\TxA}\Big)^{-1},\nonumber\\
       &\leq& \sum_{i=1}^{\TxA}\Big(1 + \frac{\Pp\Lp}{\TxA}\Big)^{-1},
\end{eqnarray}
where (\ref{eq:MSE3}) is obtained using the Schur-concavity of
$\sum_{i=1}^{\TxA}\Big(g^{-1}_i + \frac{\Pp\Lp}{\TxA}\Big)^{-1}$ in
$\vg$. From (\ref{eq:MSE2}) and (\ref{eq:MSE3}), we conclude that
\begin{eqnarray}\label{eq:MSE4}
       \text{inf}_{\vX_p \mct{\vX_p}}
\,\,\text{sup}_{\Rh}\, \text{tr}\{\Rhtilde\} &=&
\sum_{i=1}^{\TxA}\Big(1 + \frac{\Pp\Lp}{\TxA}\Big)^{-1},\nonumber
\end{eqnarray}
which can be achieved by $\vX_p \mct{\vX_p} =
\frac{\Pp\Lp}{\TxA}\vI_{\TxA}$. $\Box$

\section{Proof of Theorem 3}
\label{app:B1}

\begin{figure}[!t]
\centering\vspace{-0mm}
\includegraphics[width=0.6\columnwidth]{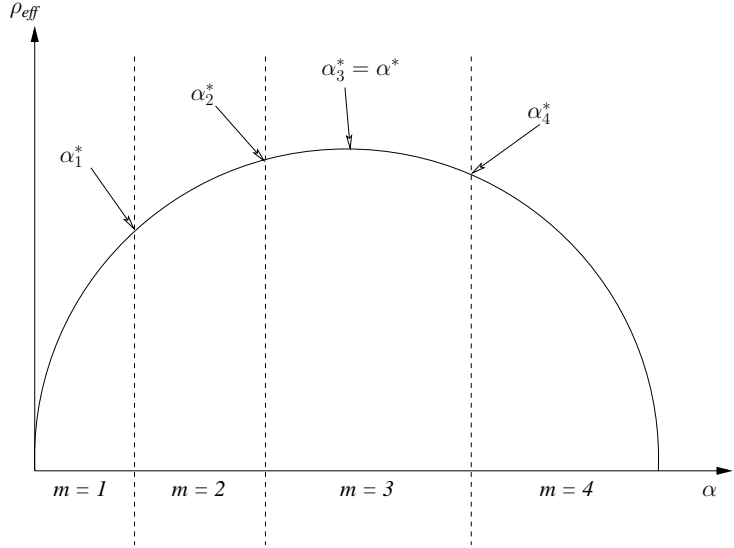}
\vspace{-6mm} \caption{A sketch example of $\SNReff$ v.s. $\alpha$.
The vertical dashed lines indicates the values of $\alpha$ at which
$m$ changes its value. $\alpha_1$, $\alpha_2$, $\alpha_3$ and
$\alpha_4$ indicate the local optimal values of $\alpha$ which gives
local maximal $\SNReff$.} \label{fig:SNReff} \vspace{-3mm}
\end{figure}

With $\shtilde = \Big(1+\frac{\Pp \Lp}{\TxA}\Big)^{-1}$, $\shhat = 1
- \shtilde$ and (\ref{eq:power}), it can be shown that $\SNReff
\triangleq \frac{\shhat\Pd}{1+\shtilde\Pd}$ is a concave function of
$\alpha \in [0,1]$. Also, $m$ is discrete and non-decreasing on
$\alpha \in [0,1]$ as the number of active eigen-channel cannot
decrease as the data transmission power increases. Here we show a
sketch plot of $\SNReff$ versus $\alpha$ in Fig.~\ref{fig:SNReff} to
visualize the proof. From (\ref{eq:MIMOInsCapLB4b}) we see that
$\Clbl$ is maximized when $\SNReff$ reaches its maximum for any
fixed $m$. Therefore, we will have $\alpha^*_1$, $\alpha^*_2$,
$\alpha^*_3$ and $\alpha^*_4$ as the local optimal points in
Fig.~\ref{fig:SNReff} which maximize $\Clbl$ in corresponding
regions of $\alpha$. From the property of water-filling solution in
(\ref{eq:waterfillingIID}), we know that $q_i$ is continuous on
$\Pd$ and hence, is continuous on $\alpha \in [0,1]$. Therefore,
$\Clbl$ in (\ref{eq:MIMOInsCapLB3}) is continuous on $\alpha \in
[0,1]$. This implies that $\Clbl$ is continuous across the
boundaries of different regions of $\alpha$, indicated by the dashed
lines in Fig.~\ref{fig:SNReff}. Consequently, the global optimal
point $\alpha^*_3 = \alpha^*$ which maximizes $\SNReff$ in
Fig.~\ref{fig:SNReff} is also the global optimal point which
maximizes $\Clbl$. It is noted that the objective function $\SNReff$
is the same as that in non-feedback systems given
in~\cite{hassibi_hochwald_03}. Therefore, the solution of $\alpha^*$
coincides with the solution for non-feedback systems given in
(\ref{eq:hassibi}). $\Box$

\section{Proof of Theorem 4}
\label{app:B2}

We let $\SNReff \triangleq \frac{\shhat\Pd}{1+\shtilde\Pd}$, $y =
\sum_{i=1}^{m} \ln \frac{\lambda_i}{m}$, and $z =
\sum_{i=1}^{m}\lambda^{-1}_i$. Then the average capacity lower bound
in (\ref{eq:MIMOaveCapLB}) can be rewritten using
(\ref{eq:MIMOInsCapLB4b}) as
\begin{eqnarray}
       \Clb &=& \frac{\Ld}{L} \frac{1}{\ln2} E_{\vlam} \{ m \ln
       (\SNReff + z) + y \}.\nonumber
\end{eqnarray}
Differentiating $\Clb$ \wrt $\Ld$ for any fixed $m$ gives
\begin{eqnarray}\label{eq:dCapdTd}
       \frac{\mathrm{d}\Clb}{\mathrm{d}\Ld} &=& \frac{1}{\ln2} \frac{m}{L} \Big(E_{\vlam} \Big\{ \ln
       (\SNReff + z) + \frac{\Ld}{\SNReff + z}\frac{\mathrm{d}\SNReff}{\mathrm{d}\Ld}+\frac{y}{m} \Big\} \Big).
\end{eqnarray}
Similar to~\cite{hassibi_hochwald_03}, we need to show that
$\frac{\mathrm{d}\Clb}{\mathrm{d}\Ld} > 0$. It can be shown that
$\Clb$ is continuous on $\Ld$ (treating $\Ld$ as a positive
real-valued variable) regardless the value of $m$. Therefore, the
value of $m$ does not cause any problem in the proof.

Here we consider the case where $\Ld > \TxA$ and omit the cases $\Ld
= \TxA$ and $\Ld < \TxA$ which can be handled similarly. Taking the
derivative of $\SNReff$ \wrt $\Ld$ with some algebraic manipulation,
we have
\begin{eqnarray}\label{eq:dSNReffdTd}
       \frac{\mathrm{d}\SNReff}{\mathrm{d}\Ld} &=&
       -\frac{\SNReff}{\Ld-\TxA}\left(1-\sqrt{\frac{\TxA(\TxA+\Pm
       L)}{\Ld(\Ld + \Pm L)}}\right).
\end{eqnarray}
Substituting (\ref{eq:dSNReffdTd}) into (\ref{eq:dCapdTd}), we get
\begin{eqnarray}\label{eq:dCapdTd2}
       \frac{\mathrm{d}\Clb}{\mathrm{d}\Ld}\!\!\! &=&\!\!\! \frac{1}{\ln2} \frac{m}{L} \left(E_{\vlam} \Big\{ \ln
       (\SNReff + z) - \frac{\SNReff}{\SNReff + z} \frac{\Ld}{\Ld-\TxA}\left(1-\sqrt{\frac{\TxA(\TxA+\Pm
       L)}{\Ld(\Ld + \Pm L)}}\right)+\frac{y}{m} \Big\}
       \right).\nonumber
\end{eqnarray}
With $\Ld > \TxA$, it can be shown that
\begin{eqnarray}\label{eq:}
    \frac{\Ld}{\Ld-\TxA}\left(1-\sqrt{\frac{\TxA(\TxA+\Pm
       L)}{\Ld(\Ld + \Pm L)}}\right) < 1.\nonumber
\end{eqnarray}
Therefore, it suffices to show that
\begin{eqnarray}\label{eq:condition1}
       E_{\vlam} \Big\{ \ln
       (\SNReff + z) - \frac{\SNReff}{\SNReff + z}+\frac{y}{m} \Big\} \geq 0.
\end{eqnarray}
Furthermore, one can show that
\begin{eqnarray}\label{eq:}
       \frac{\mathrm{d}}{\mathrm{d}\SNReff} E_{\vlam} \Big\{ \ln
       (\SNReff + z) - \frac{\SNReff}{\SNReff + z}+\frac{y}{m} \Big\}
       = \frac{\SNReff}{(\SNReff+z)^2} \geq 0\nonumber
\end{eqnarray}
for any fixed $m$. Therefore, we only need to show
(\ref{eq:condition1}) holds at $\SNReff = 0$, that is
\begin{eqnarray}\label{eq:}
       E_{\vlam} \Big\{ \ln z + \frac{y}{m} \Big\}
       &=& E_{\vlam} \Big\{ \ln \sum_{i=1}^{m}\lambda^{-1}_i + \frac{1}{m} \sum_{i=1}^{m} \ln \frac{\lambda_i}{m}
       \Big\},\nonumber\\
        &\geq& E_{\vlam} \Big\{ \frac{1}{m} \sum_{i=1}^{m} \ln \frac{\lambda^{-1}_i}{m} + \frac{1}{m} \sum_{i=1}^{m} \ln \frac{\lambda_i}{m}
       \Big\},\label{eq:condition2}\\
       &=& E_{\vlam} \Big\{ \ln \frac{\lambda^{-1}_i}{m} \frac{\lambda_i}{m} \Big\} = 0,\nonumber
\end{eqnarray}
where (\ref{eq:condition2}) is obtained using the concavity of
$\ln(\cdot)$. Therefore, we conclude that
$\frac{\mathrm{d}\Clb}{\mathrm{d}\Ld} > 0$, which implies the
training length should be kept minimum, \ie $\LpOpt = \TxA$. $\Box$

\section{Proof of Theorem 5}
\label{app:C}

For any positive definite matrix $\vA$, $\log_2|\vA|$ is increasing
in $\vA$~\cite{marshall_olkin_79}. Also, for any positive
semi-definite matrix $\vB$, $\vI + \mct{\hat{\vH}_0}\hat{\vH}_0 \vB$
is a positive definite matrix~\cite{telatar_99}. Since $\hat{\vG}
\hat{\vQ} (1+\mu^{-1}\Pd)^{-1}$ is a positive semi-definite matrix,
the capacity lower bound in (\ref{eq:MIMOInsCapLB6}) is maximized
when the diagonal entries of $\hat{\vG} \hat{\vQ}
(1+\mu^{-1}\Pd)^{-1}$ are maximized.

The $i$th non-zero diagonal entry of $\hat{\vG} \hat{\vQ}
(1+\mu^{-1}\Pd)^{-1}$ is given by
\begin{eqnarray}\label{eq:rhoeffi}
    \SNReffi &=& \frac{(g_i - \mu^{-1})\Pd}{(1+\mu^{-1}\Pd)\Lp}
     = \frac{g_i}{\Lp} \frac{\Pp \Pd + \Pd ( y - g^{-1}_i)}{\Pp + \Pd
     + y},
\end{eqnarray}
where we have used (\ref{eq:waterfillingtrain}) and let $y = \mu -
\Pp = \frac{1}{\Lp}\sum_{i=1}^{\Lp}g^{-1}_i$. Substituting $\alpha$
from (\ref{eq:power}) into (\ref{eq:rhoeffi}) with some algebraic
manipulation, we get
\begin{eqnarray}\label{eq:rhoeffi2}
    \SNReffi &=& \frac{g_i\Pm L}{\Lp (\Ld-\Lp)}
    \frac{\alpha(1-\alpha) + \alpha\frac{\Lp}{\Pm L}(y -
    g^{-1}_i)}{-\alpha + \frac{\Pm L + \Lp y}{\Pm L (1-\Lp/\Ld)}}.
\end{eqnarray}

Here we consider the case where $\Ld > \Lp$ and omit the cases $\Ld
= \Lp$ and $\Ld < \Lp$ which can be handled similarly. It can be
shown that $\SNReffi$ in (\ref{eq:rhoeffi2}) is concave in $\alpha
\in (0,1)$. Therefore, the optimal $\alpha$ occurs at
$\frac{\mathrm{d}\SNReffi}{\mathrm{d}\alpha} = 0$, which is the root
to $\alpha^2 - 2 \alpha \gamma + \gamma + \gamma z = 0$, where
$\gamma = \frac{\Pm L+ \Lp y}{\Pm L (1-\Lp/\Ld)}$ and $z =
\frac{\Lp}{\Pm L}(y-g^{-1}_i)$. It is clear that $\alpha$ depends on
$g_i$ through $z$. Therefore, there is no unique $\alpha$ which
maximizes all $\SNReffi$. However, this dependence disappears when
$\Pm L\gg \Lp y = \sum_{i=1}^{\Lp}g^{-1}_i$. Under this condition,
one can show that $\gamma \approx \frac{\Ld}{\Ld-\Lp}$ and $z
\approx 0$. And there exists a unique solution of $\alpha^*$ which
maximizes all the diagonal entries of $\hat{\vG} \hat{\vQ}
(1+\mu^{-1}\Pd)^{-1}$, given by
\begin{eqnarray}\label{eq:OptAlphaSol2}
    \alpha^* &=& \gamma - \sqrt{\gamma(\gamma-1)}, \,\,\, \text{where} \,\,\, \gamma = \frac{\Ld}{\Ld-\Lp}. \,\,\,\Box
    \nonumber
\end{eqnarray}

%\bibliographystyle{IEEEtran}
%\bibliography{IEEEabrv,paper_v4}

\end{document}